\documentclass{ws-mpla}
\usepackage[super]{cite}
\usepackage{graphicx}
\usepackage{hyperref}

\newcommand{\ANUE}{$\bar\nu_e$}
\newcommand{\Ufive}{${}^{235}{\rm U}$}
\newcommand{\Ueight}{${}^{238}{\rm U}$}
\newcommand{\PUnine}{${}^{239}{\rm Pu}$}
\newcommand{\PUone}{${}^{241}{\rm Pu}$}

\begin{document}

\markboth{David E. Jaffe}
{}

\catchline{}{}{}{}{}

\title{A Decade of Discoveries \\ by the Daya Bay Reactor Neutrino Experiment}

\author{\footnotesize David E. Jaffe\footnote{On behalf of the Daya Bay Collaboration.
}}

\address{Physics Department, Brookhaven National Laboratory\\
Upton, NY 11978, U.S.A. \\
djaffe@bnl.gov}

%

\maketitle

\pub{Received (Day Month Year)}{Revised (Day Month Year)}

\begin{abstract}
With the end of Daya Bay experimental operations in December 2020, I review the history, discoveries, measurements and impact of the Daya Bay reactor neutrino experiment in China. 

\end{abstract}

\ccode{PACS Nos.: 14.60.pq, 14.60.St}

\section{Introduction}

The Daya Bay reactor antineutrino experiment ended data-taking operations on 12 December 2020, over nine years from the start of data-taking in September 2011. 
Over this period, Daya Bay accumulated an unprecedented sample of over 6.3 million electron antineutrinos  ({\ANUE}s) detected via inverse beta decay with multiple detectors from the six nuclear reactors comprising the Daya Bay nuclear power plant near Shenzhen, China. 
Results during this time include the first definitive observation of short baseline reactor neutrino disappearance, constraints on sterile neutrinos and measurement of both the flux and spectrum of reactor \ANUE\ from the fission of \Ufive\ and \PUnine, the main fissioning isotopes of commercial reactors.

\section{Motivation for the Daya Bay experiment \label{sec:motive}}

James Chadwick's observation in 1914~\cite{ref:Pais} of a continuous $\beta$ energy spectrum was perplexing because it seemed to imply non-conservation of energy and momentum. 
Wolfgang Pauli proposed in 1930, in his famous letter~\cite{ref:Brown}, ``a desperate remedy to save...the law of conservation of energy'' by proposing a light, neutral, spin $1/2$ particle, later dubbed {the} ``neutrino" by Enrico Fermi. 
Pauli lamented in his letter that the neutrino was likely unobservable. 
In the 1950's, the challenge was overcome by Frederick Reines and Clyde Cowan when they definitively observed \ANUE\ produced by a nuclear reactor using the inverse beta-decay (IBD) reaction $\bar\nu_e + p \rightarrow e^+ + n$.~\cite{ref:Reines}

Numerous experiments since then showed that neutrinos come in three flavors, have mass and oscillate. 
The oscillations of neutrinos involve two mass-squared differences $\Delta m_{32}^2$ and $\Delta m_{21}^2 (\Delta m_{ij}^2\equiv m(\nu_i)^2 - m(\nu_j)^2)$, three mixing angles $\theta_{12},\ \theta_{23} \ {\rm and}\ \theta_{13}$, and  a CP-violating phase $\delta$. 
Around the turn of the century, it was known that $|\Delta m_{32}^2| \approx 2 \times 10^{-3}\ {\rm eV}^2$, $\theta_{23}\approx 45^\circ$, $\Delta m_{21}^2 \approx 6.9\times 10^{-5}\ {\rm eV}^2$ and $\sin^2\theta_{12} \approx 0.3$~\cite{ref:whitepaper}. 
From the CHOOZ reactor neutrino experiment~\cite{ref:CHOOZ1,ref:CHOOZ2}, it was known that $\sin^22\theta_{13} \lesssim 0.1$ at $|\Delta m_{32}^2| \approx 2 \times 10^{-3}\ {\rm eV}^2$. 

\begin{figure}[!htb]
\begin{center}
\includegraphics[height=10cm]{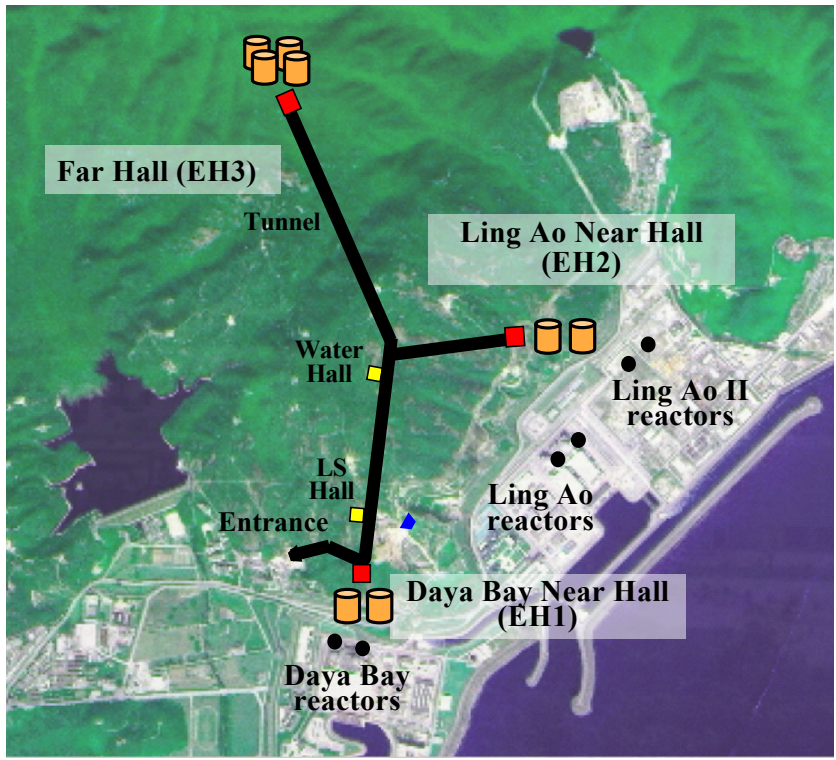}
\caption{The configuration of the Daya Bay experiment, optimized
for best sensitivity in sin$^22\theta_{13}$. Four detector modules are
deployed at the far site (EH3) and two each at each of the near sites (EH1, EH2).}
\label{fig:deployment}
\end{center}
\end{figure}

A 2004 white paper~\cite{ref:whitepaper} outlined a strategy to use multiple, identical detectors to reduce uncertainties due to reactor flux, detection efficiency and target mass which were major factors limiting CHOOZ's sensitivity to $\theta_{13}$. 
The Daya Bay collaboration was formed and proposed a reactor neutrino experiment in southern China with eight ``identical" antineutrino detectors (ADs) in three underground experimental halls (EHs) near the six reactors of the Daya Bay nuclear power plant (Figure~\ref{fig:deployment})~\cite{ref:DYBproposal}. 

A Daya Bay AD utilizes a cylindrical 3-zone design with the zones divided by thin acrylic vessels~\cite{An:2015qga}. 
The $\bar\nu_e$ target comprises 20 tons of Gd-loaded liquid scintillator (GdLS) in a 3 m high, 3 m diameter inner acrylic vessel (IAV). 
The next zone is 2{1} tons pure LS in a 4 m high, 4 m diameter outer acrylic vessel (OAV) denoted as the ``gamma catcher''. 
The outermost zone contains 192 8" photomultiplier tubes (PMTs) arranged azimuthally outside the gamma catcher submerged in 40 tons of mineral oil contained within the 5 m high, 5 m diameter stainless steel vessel. 
Reflective disks at the top and bottom of the gamma catcher create{, to first order,} an optically infinite cylinder. 
The scintillator light yield and photomultiplier coverage enabled an AD light yield of about 160 photoelectrons per MeV of deposited energy. 

The AD target scintillator contains 0.1\% Gd by mass to take advantage of Gd's large neutron capture cross-section and the emission of $\sim\!8\ {\rm MeV}$ of gamma{-ray} energy.~\cite{Yeh:2007zz, Ding:2008zzb, Yeh:2010zz, Beriguete:2014gua}
The high capture energy enhances discrimination against backgrounds from natural radioactivity with typical energies $<5\ {\rm MeV}$ and the typical capture time of {28} $\mu$s in GdLS for IBD neutrons suppresses combinatoric background. 
Selecting nGd captures efficiently selects IBDs occurring in the target. 

Each AD is equipped with three Automated Calibration Units (ACUs) that allow remote deployment of radioactive sources 
(${}^{60}{\rm Co}$, ${}^{68}{\rm Ge}$ and ${}^{241}{\rm Am}\!\!-\!\!{}^{13}{\rm C}$) and light-emitting diodes.~\cite{Liu:2013ava} 
ACUs A and B deploy sources along the central axis and near the wall of the IAV, respectively. 
ACU-C deploys sources in the gamma catcher. 
${}^{60}{\rm Co}$ produces two $\gamma$s with a total energy of 2.506 MeV used for energy calibration. 
The positron annihilation source ${}^{68}{\rm Ge}$ enables determination of the efficiency for positrons produced at IBD threshold. 
The ${}^{241}{\rm Am}\!\!-\!\!{}^{13}{\rm C}$ source is designed to produce single neutrons without $\gamma$s.~\cite{Liu:2015cra}

The ADs are deployed in 10 m deep water pools in each EH such that there is at least 2.5 m of water shielding around every AD. 
The water volume is divided into  inner and outer regions by highly reflective tyvek sheets and 8" PMTs are deployed in both regions to actively shield the ADs. Four layers of resistive plate chambers are deployed over the water pools for additional active shielding. 
Full details can be found in Refs.~\refcite{Dayabay:2014vka,Chow:2015ere}. As shown in Figure~\ref{fig:deployment}, EH1 and EH2 are near the Daya Bay and Ling Ao reactor complexes, respectively, and EH3 is the far site. 

Nuclear reactors produce $\bar\nu_e$ from beta decay of the daughters of the fissioning isotopes. 
Commercial reactors such as Daya Bay produce about $2\times10^{20}\ \bar\nu_e/{\rm s}/{\rm GW}_{th}$. 
Each of the six Daya Bay reactors operates at 2.9 ${\rm GW}_{th}$. 
The observed reactor $\bar\nu_e$ energy spectrum is the result of the convolution of the steeply falling $\bar\nu_e$ spectrum from beta decay and the rising IBD cross-section with a threshold of $\sim\!1.8$ MeV. The spectrum peaks around 3 to 4 MeV and drops to zero at about 8 MeV.~\cite{Bemporad:2001qy} 

Disappearance of \ANUE\ is governed by 
\begin{equation}
P(\bar\nu_e\!\to\!\bar\nu_e) = 
\!1 - 
\sin^22\theta_{13} \sin^2\left(\Delta m_{ee}^2\frac{L}{4E}\right) - 
\sin^22\theta_{12}\cos^4\theta_{13}\sin^2\left(\Delta m_{21}^2\frac{L}{4E}\right)
\label{eqn:anuedis}
\end{equation}
where $\sin^2(\Delta m_{ee}^2\frac{L}{4E})  \equiv \cos^2\theta_{12}\sin^2(\Delta m_{31}^2\frac{L}{4E}) + \sin^2\theta_{12}\sin^2(\Delta m_{32}^2\frac{L}{4E})$ is valid for the \ANUE\ energy ($E$) and baseline ($L$) ranges of Daya Bay~\cite{ref:dmee}. 
Note that $|\Delta m_{ee}^2|$ is related to the actual mass splitting 
\begin{equation}
|\Delta m_{ee}^2| \approx |\Delta m_{32}^2| \pm 5\times10^{-5}\ {\rm eV}^2
\label{eqn:dmee}
\end{equation}
 where the $+$($-$) sign corresponds the normal(inverted) hierarchy. 
The $\sin^22\theta_{12}$ term is negligible for Daya Bay baselines.


The two competing, contemporaneous experiments, Double Chooz~\cite{Ardellier:2006mn} in France and RENO~\cite{Ahn:2010vy} in South Korea, are compared with Daya Bay in  Table~\ref{tab:exptcomp}. 
Daya Bay has advantages in signal yield thanks to significantly greater target mass and comparable or greater reactor power and reduced cosmogenic background due to greater overburden. 

\begin{table}[h]
\tbl{
Comparison of the reactor power, target mass{, overburden and far site baseline} of the Double Chooz, RENO and Daya Bay experiments. 
{ The parameters of the upcoming JUNO experiment are also provided.~\cite{Abusleme:2021zrw}}
}
{\begin{tabular}{lcccccc}
\hline
                     & Reactor                   & \multicolumn{2}{c}{Target (tons)} & \multicolumn{2}{c}{Overburden (m.w.e)} & {Baseline (km)} \\
Experiment   & power (GW$_{th}$) & Near & Far   				& Near & Far 						 & { Far}\\
\hline
Double Chooz 	    & 8.6   & 8 & 8 	    & 120 & 300 & {1.05}\\
RENO              	    &16.5  &16&16	    & 120 & 450 & {1.38}\\
Daya Bay         	    &17.4  &80&80	    & 250 & 860 & {1.65}\\
{ JUNO }&{ 26.6 }&{ ---} &{ 20000 }&{ --- }&{ 1800 }&{ 53}\\
\hline
\end{tabular}
\label{tab:exptcomp} }
\end{table}%

\section{Observation of non-zero $\theta_{13}$} 

In 2011, long baseline accelerator neutrino experiments MINOS~\cite{Adamson:2011qu} and T2K~\cite{Abe:2011sj} had hints of non-zero $\theta_{13}$ from putative $\nu_e$ appearance in $\nu_\mu$ beams. 
Additional indications of non-zero $\theta_{13}$ included Double Chooz~\cite{Abe:2011fz} results using only their far detector and an analysis using the combination of solar neutrino and KamLAND results~\cite{Fogli:2011qn}. 
These results motivated Daya Bay to re-evaluate the deployment and commissioning plan in mid-2011. 
At the time, it was known that RENO would soon start data-taking and that T2K might resume data-taking before the end of 2011. 
Daya Bay's altered plan called for  a ``2-1-3" deployment of two ADs in EH1, one AD in EH2 and three ADs in EH3 with a goal of a first  publication in March 2012. 

EH1 operation began 23 September 2011, EH2 operation in mid-October and EH3 operation on 24 December 2011. 
With the initial three months of EH1 operation with two ADs from 23 Sept. to 23 Dec. 2011, Daya Bay demonstrated that the two ADs were indeed ``functionally identical" with measured inter-detector systematic uncertainties better than required~\cite{DayaBay:2012aa}. 

The Daya Bay collaboration held an analysis workshop at the start of 2012 at the Daya Bay site in southern China. 
Preliminary analysis results from the first five days of ``2-1-3" data-taking were presented at the workshop on 3 January 2012. 
These results indicated a deficit in IBD events in EH3 with respect to the prediction based on the near sites, EH1 and EH2, of roughly three standard deviations. 
Needless to say, this surprising indication of a large $\theta_{13}$ was quite a motivation to the collaboration to re-double the effort to meet the March 2012 goal. 

\begin{figure}[!htb]
\begin{center}
\includegraphics[height=5cm]{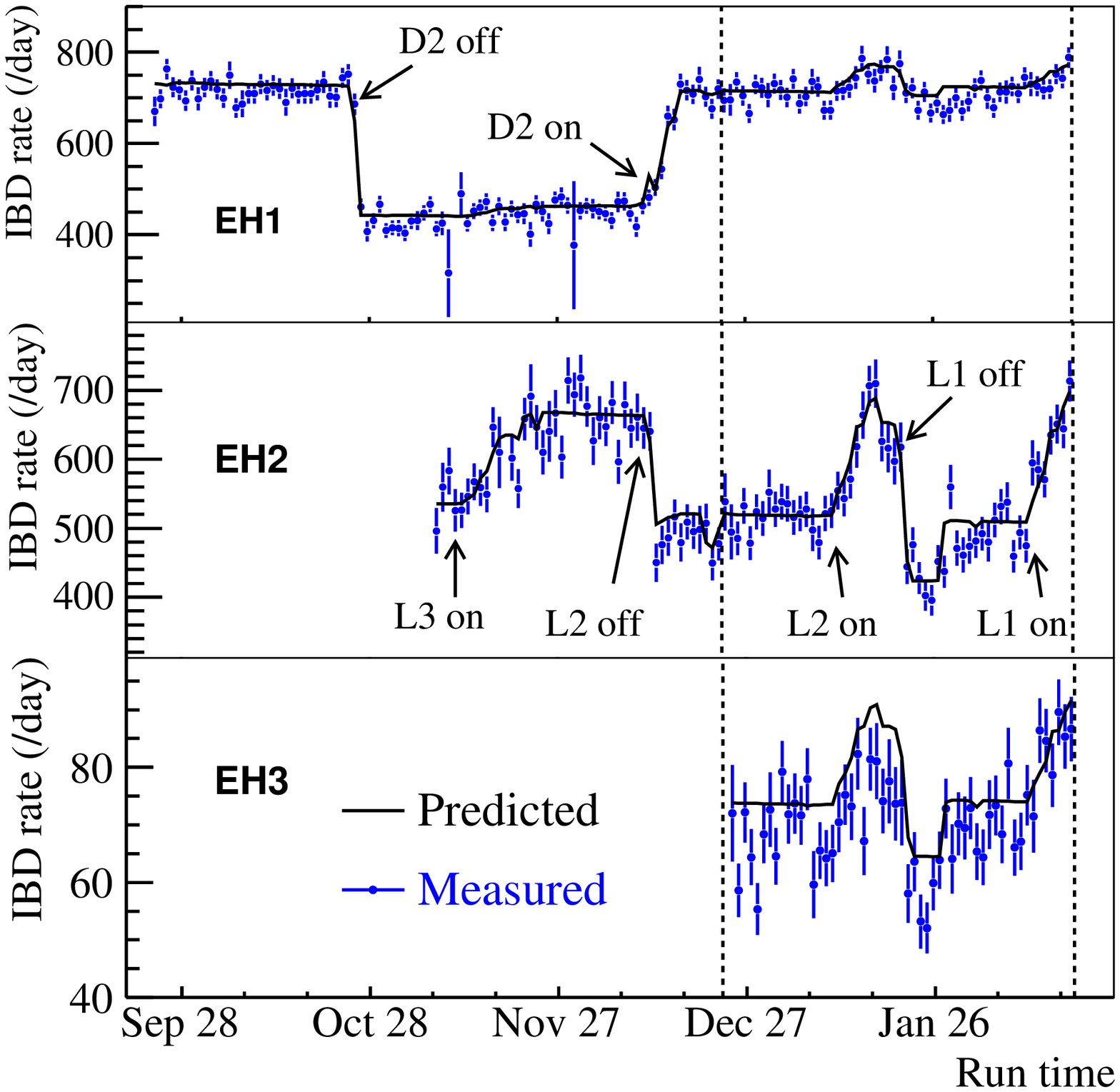}
\includegraphics[height=5cm]{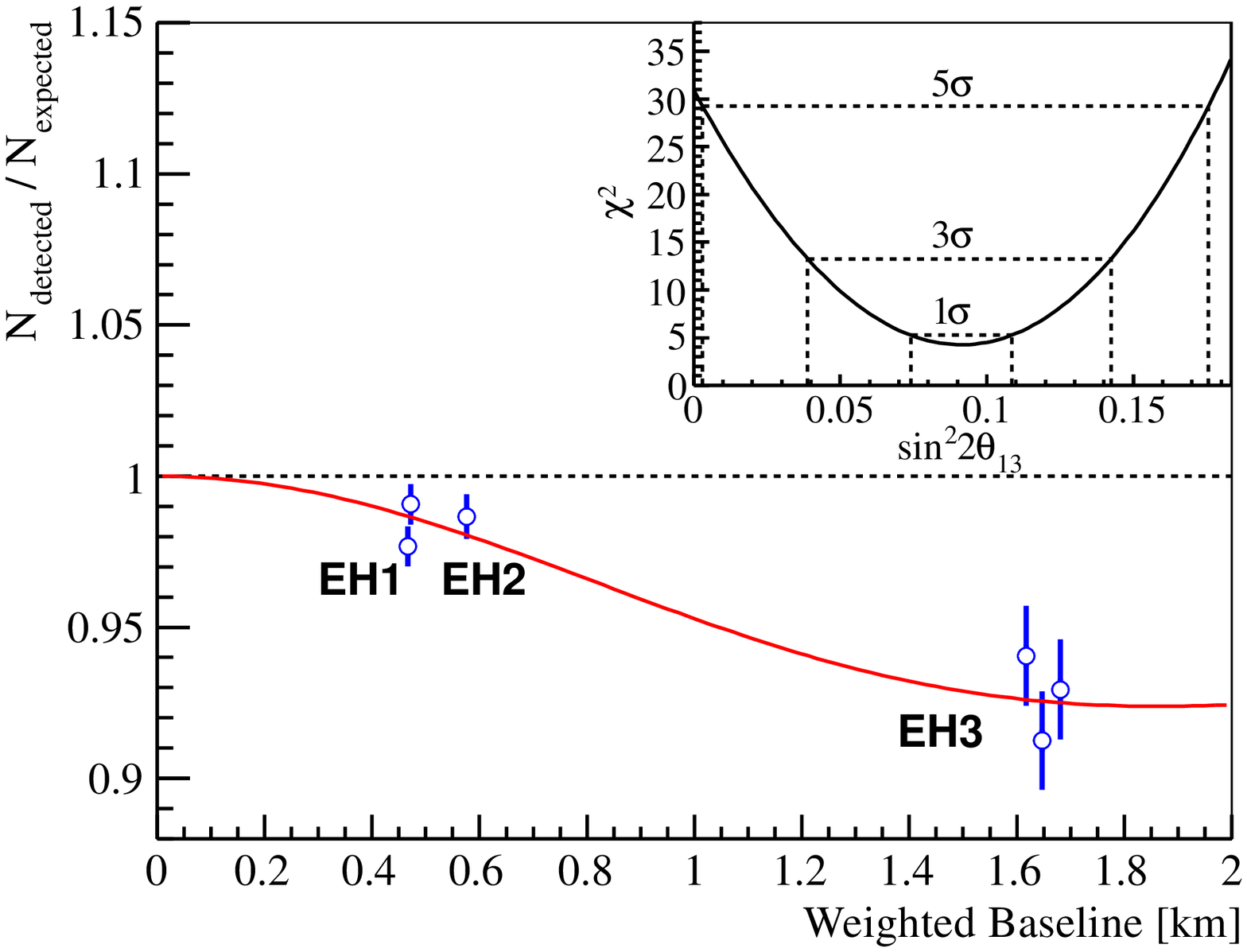}
\caption{The left figure shows the observed daily IBD rate in the three experimental halls compared to the expection assuming no oscillations using information from the reactor company. The states of the Daya Bay and Ling Ao reactors are indicated by  ``D" and ``L", respectively. The right figure shows the total rate in each AD relative to expectation as a function of the weighted baseline. The expectation is corrected with the best-fit normalization parameter. The smooth curve shows the best-fit oscillation survival probability. The inset shows $\chi^2$ as a function of sin$^22\theta_{13}$. 
{Figures from Ref.~\protect\refcite{An:2012eh}.} 
}
\label{fig:firstres}
\end{center}
\end{figure}

The initial results using 55 days of data confirmed the indications of the first five days of data and yielded the first definitive observation of non-zero $\theta_{13}$ at 5.2 standard deviations.~\cite{An:2012eh} 
Figure~\ref{fig:firstres} shows that the observed IBD candidates at each site tracked the expectation based on the reactor power information provided by the reactor company as well as the clear deficit in EH3 rate with respect to EH1 and EH2. 
Interpretation of the observation using Eqn.~\ref{eqn:anuedis} results in $\sin^22\theta_{13} = 0.092\pm0.016({\rm stat.})\pm0.005({\rm syst.})$. 

As noted in Ref.~\refcite{Conrad:2012zzb}, a large value of $\theta_{13}$ enables the pursuit of the mass hierarchy and CP violation in appearance experiments that should take data soon. It also reinforces the quantitative differences between the quark and lepton mixing matrices.

A coordinated series of seminars around the world at collaborating Daya Bay institutions  announced the results on 8 March 2012 with the first (earliest) seminar at IHEP in Beijing. 
RENO confirmed the results about one month later~\cite{Ahn:2012nd}. 
The results were hailed as one of the top ten breakthroughs of 2012 by Science~\cite{ref:science}. 
Daya Bay, along with the SuperK, {T2K,}  KamLAND and SNO experiments, were awarded the 2016 Breakthrough Prize in Fundamental Physics in recognition of investigation of neutrino oscillations\cite{ref:breakthroughprize}. 

An improved measurement $\sin^22\theta_{13} = 0.089\pm0.010({\rm stat.})\pm0.005({\rm syst.})$ was reported by Daya Bay at the Neutrino 2012 conference~\cite{ref:neutrino2012} with about 2.5 times more data.~\cite{An:2012bu} 
Operations were paused after 217 days of data in the ``2-1-3'' configuration to install the final two ADs in the summer of 2012. 
A manual calibration system, capable of placing radioactive sources  throughout the target volume, was deployed during summer 2012.~\cite{Huang:2013uxa} 
{
Daya Bay operated in the ``2-2-4'' mode until January 2017 when an AD in EH1 was repurposed for JUNO LS development.~\cite{Abusleme:2020bbm} 
The experiment operated in a ``1-2-4'' mode until the end of operations in December 2020. 
}

\section{Selection of IBD candidates}

As noted in Section~\ref{sec:motive}, the inverse beta-decay reaction $\bar\nu_e + p \rightarrow e^+ + n$ provides a pair of time-correlated energy deposits. 
The positron ionizes and annihilates, producing a prompt energy deposit in the 1 to 8 MeV range. 
In the GdLS volume, the neutron captures with a characteristic time of $\sim\!28\ \mu{\rm s}$ with 84\% of captures on either ${}^{157}{\rm Gd}$ or ${}^{155}{\rm Gd}$ with relative probabilities of 81.5\% and 18.5\%, respectively. 
The $\gamma$ energy released per capture is 7.95 MeV for ${}^{157}{\rm Gd}$ and 8.54 MeV for ${}^{155}{\rm Gd}$. 
The remainder of captures (16\%) occur on hydrogen which releases a single 2.22 MeV $\gamma$. 

These characteristics allow a straightforward selection of IBD candidates using nGd captures in the Daya Bay ADs.~\cite{An:2016ses} 

\begin{enumerate}
\item \label{it:pe}The prompt energy is required to be in the range 0.7 to 12 MeV, ensuring nearly 100\% efficiency for the prompt candidate. 
\item \label{it:de}The delayed energy must be in the range 6 to 12 MeV which enables high efficiency for the nGd capture $\gamma$s.
\item \label{it:dt}The delayed candidate is required to be later than the prompt candidate by $\Delta t \in (1,200)\ \mu{\rm s}$. 
Candidates with $\Delta t < 1\ \mu{\rm s}$ are rejected since they would fall within a single triggered readout of the detector. 
\item \label{it:multv}Additional criteria, dubbed the multiplicity veto, are applied to reject occurrences of more than two signals close in time that result in confusion as to which pair is the result of an \ANUE\ interaction. 
The multiplicity veto required only one $(0.7,12)\ {\rm MeV}$ signal in the 400 $\mu$s before the delayed candidate and no $(6,12)\ {\rm MeV}$ signal in  the 200 $\mu$s after the delayed candidate. 
\item \label{it:muonv} Cosmogenic backgrounds are suppressed by rejecting candidates that occur within $(-2, 600)\ \mu{\rm s}$ of a signal in either the inner or outer regions of the water pool, or 
\item \label{it:adv} candidates that occur within $(0,1400)\ \mu{\rm s}$ of a signal with  $>\!3000$ photoelectrons in the AD, or 
\item \label{it:adsh} within $(0, 0.4)\ {\rm s}$ of a signal with $>\!300,\!000$ photoelectrons in the AD.  
\end{enumerate}
Daya Bay typically had two sets of selection criteria devised by two largely independent analysis teams.~\cite{An:2016ses} 
Criteria (\ref{it:pe})--(\ref{it:dt}) were used by both Selection A and B. 
Selection B in  Ref.~\refcite{An:2016ses} used the criteria (\ref{it:multv})--(\ref{it:adsh}). 
Selection A used analogous  criteria targetting the same backgrounds. 

Both Selections rejected instrumental background, caused by light emission from PMTs, by exploiting the sharing and pattern of light among the PMTs. 
These criteria { were} determined to have a $99.98\pm0.01\%$ efficiency for IBD events with negligible background. 
See Ref.~\refcite{An:2016ses} for details. 

Two uncorrelated signals can be selected as an IBD candidate. 
This background is dubbed ``accidentals'' and can be precisely estimated by pairing uncorrelated signals satisfying the prompt and delayed criteria. 

\begin{figure}[!htb]
\begin{center}
\includegraphics[height=9cm]{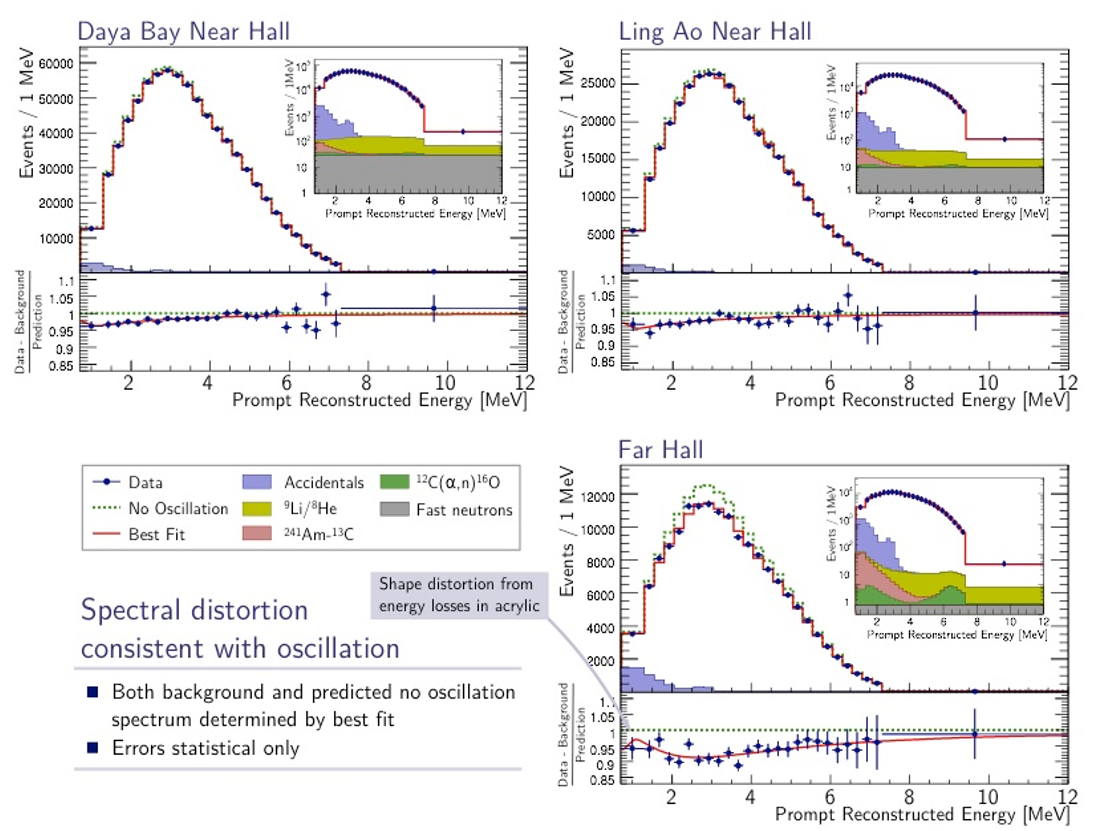}
\caption{The measured and best-fit spectra in the three experimental halls.~\cite{An:2013zwz} 
The shaded histograms show the estimated backgrounds with the inset showing the shape and level of each background.}
\label{fig:3H}
\end{center}
\end{figure}

There were four significant correlated backgrounds: 
\begin{enumerate}
\item Fast neutrons: Energetic neutrons can be generated by muons passing through or near the AD. 
A nuclear collision of a fast neutron in the scintillator { producing a recoiling proton} can mimic a prompt signal and the subsequent capture is identical to the delayed signal. 
Criteria (\ref{it:muonv})--(\ref{it:adsh}) suppress this background to $\lesssim0.1\%$ {(all percentages are relative to signal)}. 
\item $\beta$-n decays: Muons can occasionally interact {with the liquid scintillator} and produce unstable isotopes ${}^9{\rm Li}$ and ${}^8{\rm He}$ 
which can $\beta$-decay while simultaneously producing a neutron. 
Criteria (\ref{it:adv}) and (\ref{it:adsh}) suppress this background to the 0.3\% to 0.4\% level, depending on the experimental hall. 
\item ${}^{241}{\rm Am}\!\!-\!\!{}^{13}{\rm C}$ neutron sources: Neutrons from the ${}^{241}{\rm Am}\!\!-\!\!{}^{13}{\rm C}$ calibration sources parked in the ACUs can produce  $\gamma$s correlated in time that enter the active AD volume. 
This background occurs at $\lesssim0.1\%$. 
\item $(\alpha,n)$ interactions: Natural radioactivity within the detector produces $\alpha$s that can interact with stable nuclei to produce neutrons with ${}^{13}{\rm C}(\alpha,n)^{16}{\rm O}$ interactions being the most abundant. 
Protons from neutron recoil or ${}^{16}{\rm O}^*$ de-excitation $\gamma$s provide the prompt candidate. 
Contamination from this background is $\lesssim 0.07\%$.
\end{enumerate}
The magnitude and energy spectra of the IBD signal and backgrounds are shown in Figure~\ref{fig:3H} with the results from Ref.~\refcite{An:2013zwz} based on the full six-AD running period. 
The background-to-signal ratio is $\sim\!2\%$ and $\sim\!5\%$ in the near and far sites, respectively, demonstrating the effects of reactor power, total detector mass and overburden shown in Table~\ref{tab:exptcomp}. 

\section{Precision neutrino oscillation measurements \label{sec:osc}}

Subsequent to establishing non-zero $\theta_{13}$, Daya Bay has made a series of measurements of increasing precision of the neutrino oscillation parameters in Eqn.~\ref{eqn:anuedis}, namely $\sin^22\theta_{13}$ and $\Delta m_{ee}^2$. 
Measurement of the latter requires precise knowledge of the detector energy response to IBD positrons since the $\bar\nu_e$ energy is given by $E_{\bar\nu_e} \approx E_p + \bar{E}_n  + 0.78\ {\rm MeV}$ where $E_p$ is the prompt energy (both the positron kinetic energy and the annihilation energy) and $\bar{E}_n\sim10\ {\rm keV}$ is the average neutron recoil energy. 
The non-linearity in the AD energy response is dominated by two effects: the particle-type dependency of the scintillator light yield and charge-dependence of the PMT readout electronics. 
The light yield non-linearity is related to intrinsic scintillator quenching and Cherenkov light emission. 
The electronics non-linearity results from the interaction  of the scintillation light time profile and the charge-collection characteristics of the electronics.
Each effect is at the 10\% level. 

The energy response model has evolved as the details of the response have become better understood. 
Model parameters were {determined}  from gamma{-ray} lines in AD calibration data {obtained} from deployed and naturally occurring sources as well as the continuous $\beta$-decay spectrum of ${}^{12}{\rm B}$ from muon spallation in the GdLS volume. 
The initial model attained an overall uncertainty of about 1.5\% on $E_p$ and differed by $<\!0.3\%$ among the 6 ADs which had negligible impact on the initial $\Delta m_{ee}^2$ measurement.~\cite{An:2013zwz} 
The subsequent model reduced the absolute uncertainty to $<\!1\%$ for $E_p > 2\ {\rm MeV}$ and the uncorrelated uncertainty between ADs to 0.2\%.~\cite{An:2015rpe} 
The most recent  model achieves an absolute uncertainty of less than 0.5\% for $E_p>2\ {\rm MeV}$ and  0.1\% among ADs by taking into account subtle optical effects of the source enclosures and deployment system.~\cite{Adey:2019zfo} 
It also takes advantage of a dual readout system installed on one AD in EH1 in December of 2015 where a full FADC readout system acquired data simultaneously with the original readout system.~\cite{Huang:2017abb} 

The latest oscillation measurements from Daya Bay are based on nearly four million IBD candidates from 1958 days of data and yield 
$\sin^22\theta_{13} = 0.0856\pm0.0029$ and 
$\Delta m_{32}^2 = (2.471^{+0.068}_{-0.070})\times 10^{-3}\ {\rm eV}^2 $ assuming the normal {mass} hierarchy and 
$\Delta m_{32}^2 =-(2.575^{+0.068}_{-0.070})\times 10^{-3}\ {\rm eV}^2 $ assuming the inverted hierarchy.~\cite{Adey:2018zwh} 
The measurement of $\Delta m_{32}^2$ from electron antineutrino disappearance is compared to that from muon neutrino disappearance in Table~\ref{tab:dm}. The precision of the Daya Bay $\Delta m_{32}^2$ result is comparable or superior to all currently published results. 
The consistency of the results provides strong confirmation of the 3-flavor model of neutrino mixing. 
\begin{table}[h]
\tbl{Latest measurements of $\sin^22\theta_{13}$ from reactor experiments and $\Delta m_{32}^2$ under the normal (NH) and inverted (IH) hierarchy assumptions from muon and electron neutrino disappearance measurements. }
{\begin{tabular}{lcccr}
\hline
                   &  					& \multicolumn{2}{c}{$\Delta m_{32}^2$ ($10^{-3}\ {\rm eV}^2$)} &  \\
Experiment &$\sin^22\theta_{13}$  &  NH    				& IH 			& Reference \\
\hline
Daya Bay & $0.0856\pm0.0029$	& $2.471^{+0.068}_{-0.070}$	& $-2.575^{+0.068}_{-0.070}$  &\refcite{Adey:2018zwh} \\
RENO	& $0.0896\pm0.0068$	& $2.63\pm0.14$			& $-2.73\pm0.14$  &\refcite{Bak:2018ydk} \\
Double Chooz & $0.105\pm0.014$ &      				&				& \refcite{DoubleChooz:2019qbj}\\
NOVA 	& 					& $2.48^{+0.11}_{-0.06}$		& $-2.54^{+0.06}_{-0.11}$  &\refcite{Acero:2019ksn} \\
SuperK     & 					& $2.50^{+0.13}_{-0.20}$		& $-2.58^{+0.08}_{-0.37}$  &\refcite{Abe:2017aap} \\
T2K		& 					& $2.463^{+0.071}_{-0.070}$	& $-2.507\pm 0.070 $  &\refcite{Abe:2018wpn} \\
\hline

\end{tabular}
\label{tab:dm} }
\end{table}%

With the full data sample of over six million IBD candidates registered by nGd capture, the relative precision on $\sin^22\theta_{13}$ and $|\Delta m_{32}^2|$ is expected to improve from 3.4\% and 2.8\%, respectively, to 2.7\% and 2.1\%. 
The ultimate Daya Bay $|\Delta m_{32}^2|$ precision is likely to be exceeded { by} JUNO~\cite{Abusleme:2021zrw} and experiments in long baseline muon neutrino beams in the next decade. 
The precision of Daya Bay's final $\sin^22\theta_{13}$ measurement is unlikely to be surpassed for decades. 

\section{Sterile neutrinos}

In 2011, an improved prediction of reactor antineutrino spectra was calculated for the four main fissioning isotopes \Ufive, \PUnine, \Ueight\ and \PUone.~\cite{Mueller:2011nm} 
Notably, this new calculation predicted about 3\% higher \ANUE\ flux from commercial reactors. 
The disagreement between the prediction and the available measurements of \ANUE\ flux at 98.6\% C.L. was dubbed the ``Reactor Antineutrino Anomaly'' (RAA).~\cite{Mention:2011rk} 
One explanation of the RAA postulated the existence of a fourth, non-standard neutrino with $|\Delta m^2_{\rm new}| \sim {\cal O}(1\ {\rm eV}^2)$ and $\sin^22\theta_{\rm new} \sim {\cal O}(0.10)$. 
An independent calculation verified the improved prediction.~\cite{Huber:2011wv} 
The combination of the two calculations, referred to as the ``Huber-Mueller'' model, became the de facto standard for calculation of reactor antineutrino spectra. 

Reactor antineutrino experiments would be sensitive to such a new neutrino state if the disappearance of \ANUE\ deviated from Eqn.~\ref{eqn:anuedis}. 
For short baselines, such as those available at Daya Bay, \ANUE\ disappearance would be described by 
\begin{equation}
P(\bar\nu_e\to\bar\nu_e)
 \approx 
1 - 
\sin^22\theta_{13} \sin^2\frac{\Delta m^2_{31} L}{4E} -
\sin^22\theta_{14} \sin^2\frac{\Delta m^2_{41} L}{4E} 
\label{eqn:sterile}
\end{equation}
if there were a fourth neutrino state. 
This fourth neutrino is referred to as a ``sterile'' neutrino because it { has nearly zero interaction amplitude with}
the $Z$ and $W^\pm$ bosons of the Standard Model.{~\cite{Naumov:2019kwm}}

\begin{figure}[!htb]
\begin{center}
\includegraphics[height=5cm]{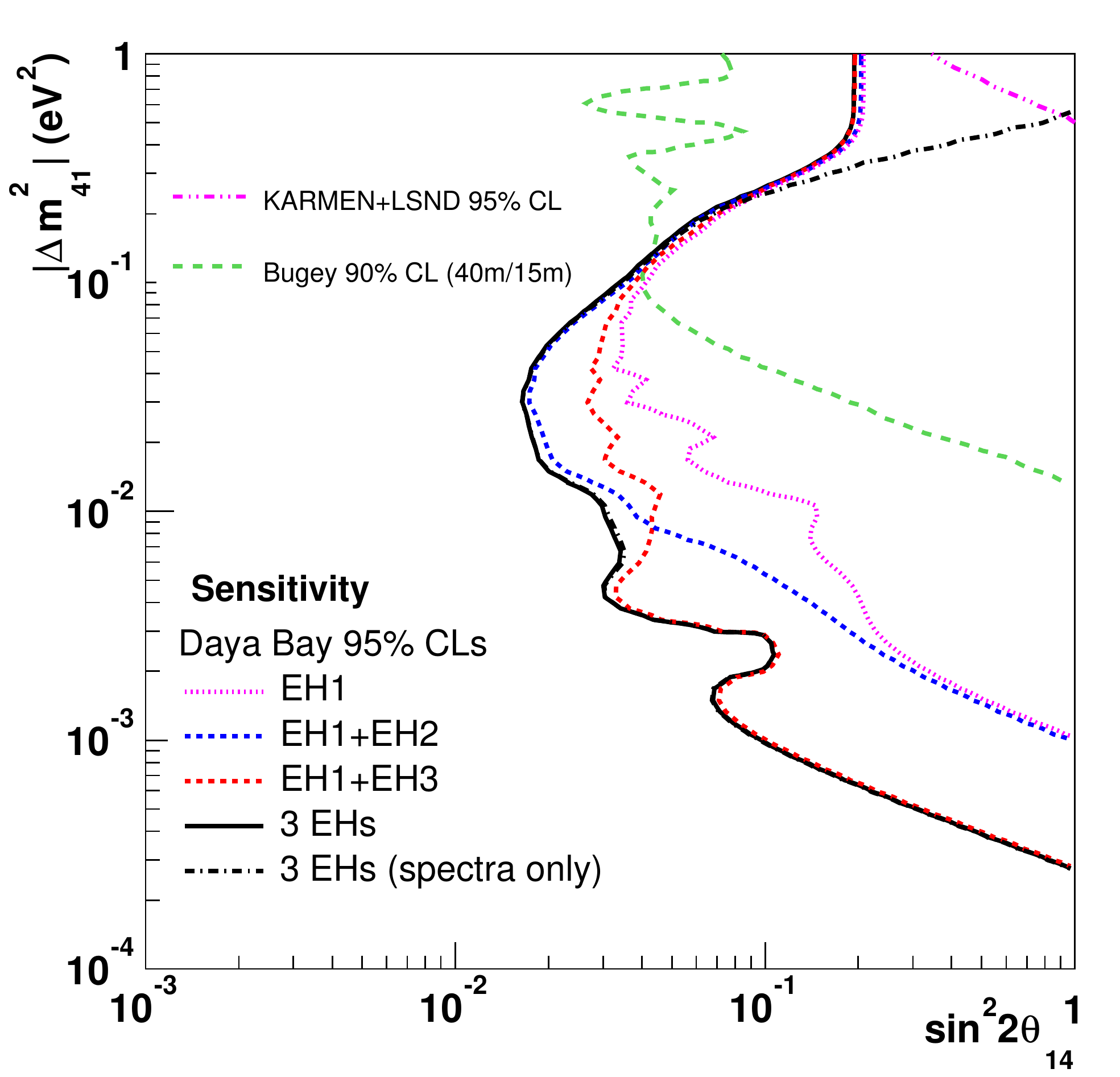}
\includegraphics[height=5cm]{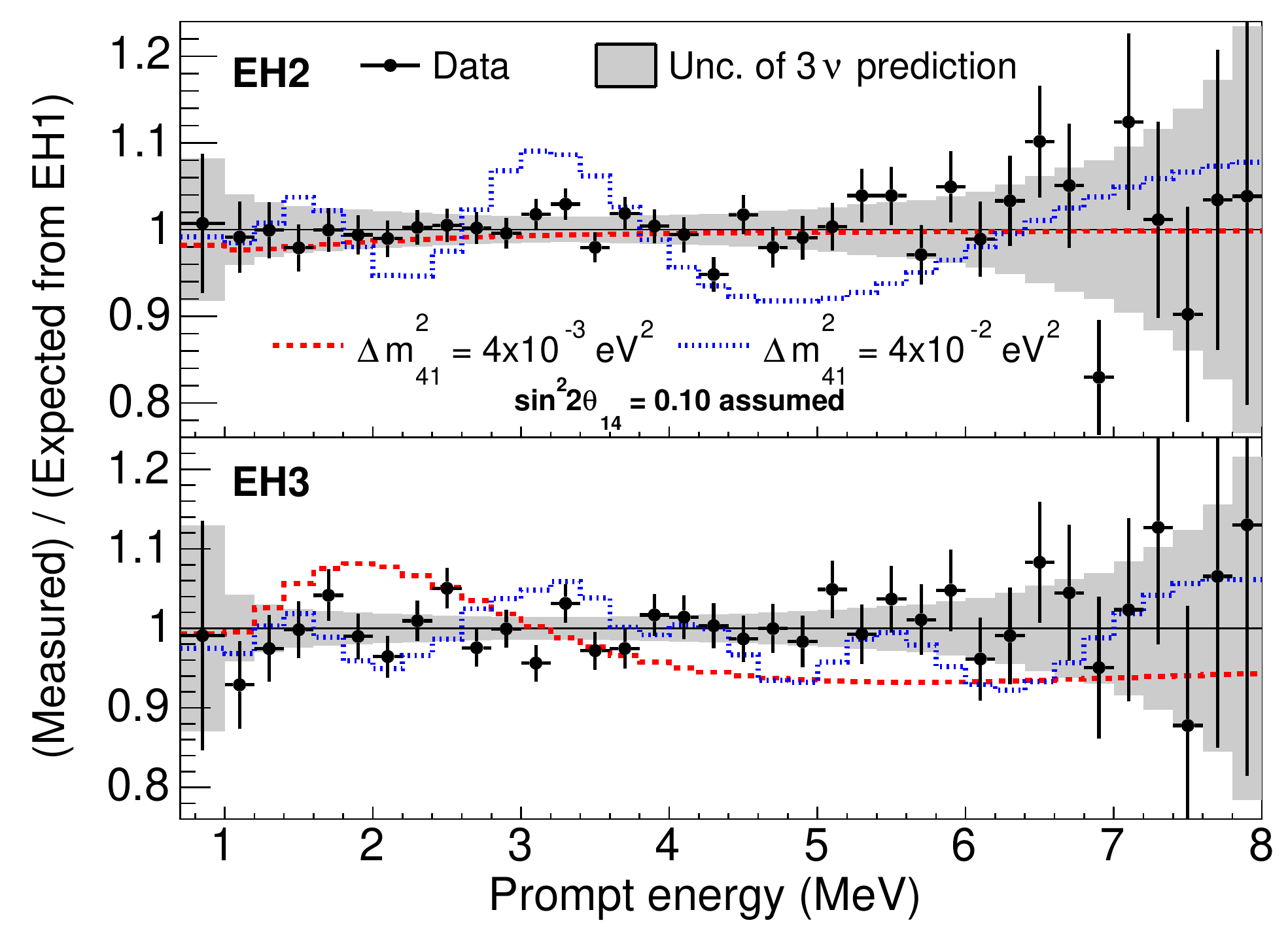}
\caption{The left figure illustrates the advantages of the multiple baselines of Daya Bay as described in the text, and
the green and magenta double-dot-single-dashed lines represent the limits from the Bugey~\cite{Declais:1994su} and the combination of KARMEN and LSND~\cite{Conrad:2011ce}, respectively. 
The upper and lower panels of the right figure shows the spectral data in EH2 and EH3 compared to the expectation from the spectrum observed in EH1, respectively,  to two hypothesized  $|\Delta m^2_{41}|$ values. While the dashed red line for $|\Delta m^2_{41}| = 0.004\ {\rm eV}^2$ appears consistent with the EH2 data, it is clearly inconsistent with the EH3 data. 
Conversely, while the dashed blue line representing $|\Delta m^2_{41}| = 0.04\ {\rm eV}^2$ provides a decent description of the EH3 data, it  obviously disagrees with the EH2 data. 
{ Figures from Ref.~\protect\refcite{An:2014bik}.}
}
\label{fig:sterile}
\end{center}
\end{figure}

The multiple baselines of Daya Bay provide sensitivity to the $L/E$ region of 50 to 200 m/MeV, corresponding to $|\Delta m^2_{41}|$ in the 0.001 to 0.1 ${\rm eV}^2$ region. 
As with the $\sin^22\theta_{13}$ measurement, the comparison of different EHs removes sensitivity to the knowledge of the absolute {anti}neutrino flux. 
The combination of EH1 and EH3 {better} constrains the $|\Delta m^2_{41}|\!<\! 0.01\ {\rm eV}^2$ region while EH1 and EH2 {better} constrains the 
$0.01 \!<\! |\Delta m^2_{41}|\!<\! 0.1\!\ {\rm eV}^2$ region as shown in Figure~\ref{fig:sterile}. 
Sensitivity to $| \Delta m^2_{41}|\!>\! 0.1\ {\rm eV}^2$, near the region favored by the RAA, is based on the comparison of the measurement of the absolute flux in the near sites, primarily EH1, with the prediction. 
The uncertainty on the absolute flux measurement is conservatively set at 5\% as illustrated in Figure~\ref{fig:sterile}. 
No evidence for a sterile neutrino from \ANUE\ disappearance was found in Daya Bay's initial search~\cite{An:2014bik}  or subsequently.~\cite{An:2016luf,Adamson:2020jvo}

The LSND experiment reported evidence~\cite{Aguilar:2001ty} for $\bar\nu_\mu\!\to\!\bar\nu_e$ appearance for $L/E \sim {\cal O}(1)\ {\rm m}/{\rm MeV}$. 
MiniBooNE, operating in an accelerator muon neutrino beam with the same $L/E$ as LSND, has observed low-energy excesses {attributed to} $\nu_e$ and $\bar\nu_e$~\cite{Aguilar-Arevalo:2012fmn} that show consistency with LSND, but are not definitive. 
The standard 3-neutrino mixing model can be extended by adding one or more sterile neutrinos to describe muon neutrino to electron neutrino transitions consistent with  {the} LSND and MiniBooNE observations.~\cite{Zyla:2020zbs} 
{ In this case the appearance probability is given by 
$P(\nu_\mu\! \to \!\nu_e) \approx \sin^22\theta_{\mu e}  \sin^2 \frac{\Delta m_{41}^2 L}{4E}$.}  
The relation between appearance and disappearance probabilities yields
{
\begin{equation}
4\sin^22\theta_{\mu e} \approx \sin^22\theta_{14} \sin^22\theta_{24}
\label{eqn:app}
\end{equation}
\noindent in the limit of small mixing with one sterile neutrino. 
}

\begin{figure}[!htb]
\begin{center}
\includegraphics[height=5.9cm]{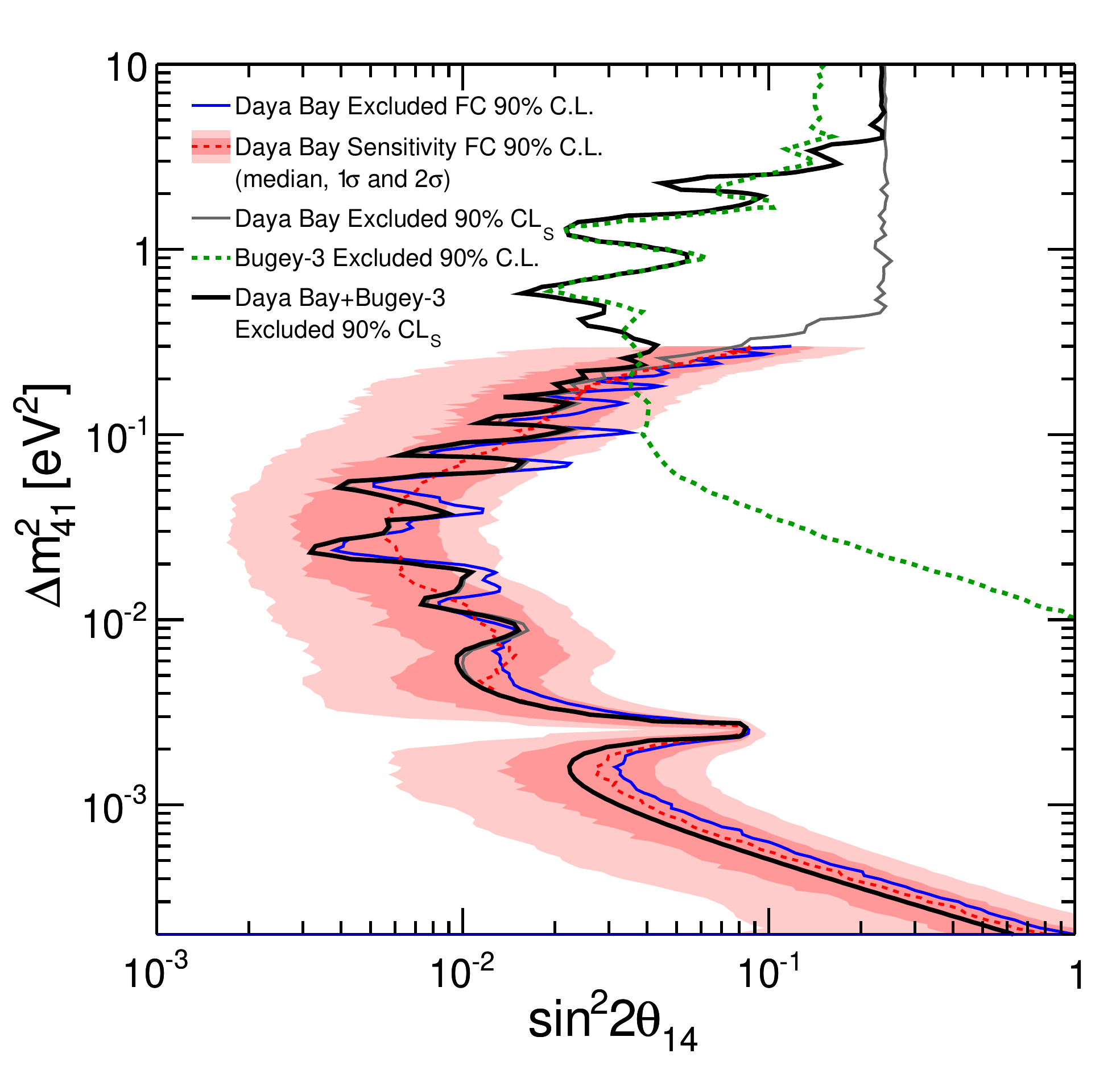}
\includegraphics[height=5.9cm]{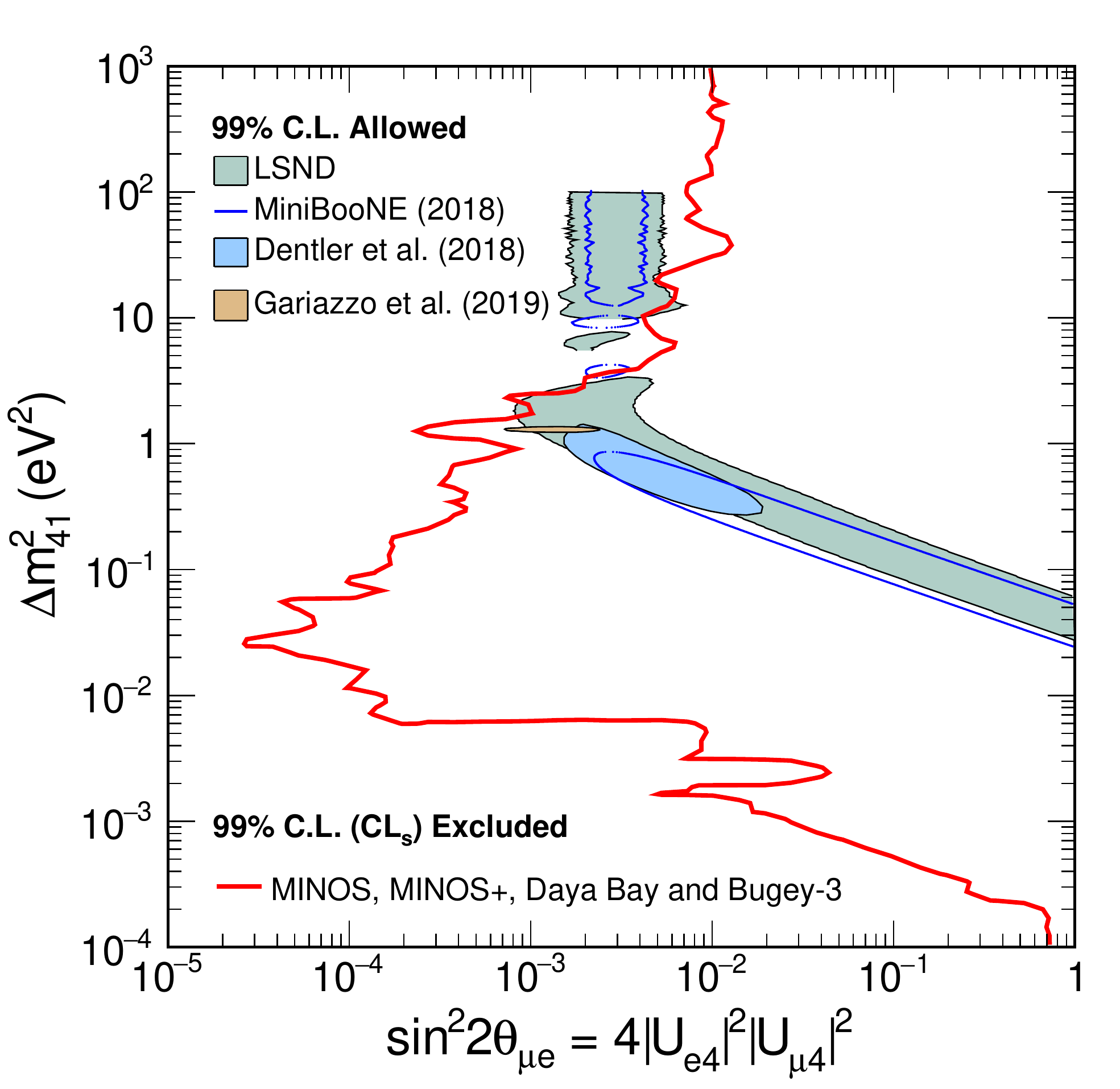}
\caption{The left figure shows the most recent limits on $\sin^22\theta_{14}$ as a function $|\Delta m_{41}^2|$.~\cite{Adamson:2020jvo}.
 The right figure shows the combined limits from Daya Bay, Bugey, MINOS, and MINOS+ data 
 on $\sin^22\theta_{e\mu}$ vs $|\Delta m_{41}^2|$.~\cite{Adamson:2020jvo}
 {Figures from Ref.\protect\refcite{Adamson:2020jvo}.}
 }
\label{fig:sterile_combo}
\end{center}
\end{figure}

Daya Bay has collaborated with MINOS~\cite{Adamson:2016jku} and MINOS+~\cite{Adamson:2020jvo} to combine \ANUE\ disappearance results with the muon neutrino disappearance measurements to constrain electron neutrino appearance from muon neutrino sources. 
The resulting limit on $\sin^22\theta_{\mu e}$ as a function of $\Delta m_{41}^2 $, taken from Ref.~\refcite{Adamson:2020jvo}, is shown in Figure~\ref{fig:sterile_combo}. 
As stated in Ref.~\refcite{Adamson:2020jvo}, ``...the LSND and MiniBooNE 99\% C.L. allowed regions are excluded at 99\% CLs for $\Delta m_{41}^2 \!<\! 1.6\ {\rm eV}^2$.'' 
The results further reduce the viability of a sterile neutrino explanation of putative electron neutrino appearance and null disappearance  results. 

\section{Measuring the reactor {\ANUE\ }flux and spectrum}

The RENO experiment {presented} preliminary results at the Neutrino 2014 conference~\cite{ref:Neutrino2014} that showed an excess of IBD candidates near 5 MeV in prompt energy with respect to the Huber-Mueller model.~\cite{Seo:fiveMeV} 
When the observed IBD {yield} is normalized to the prediction for 0.5 to 4.0 MeV, then the 5 MeV excess corresponds to $\sim\! 2\%$ of the total expected \ANUE\ flux. 
In 2016, Daya Bay published a measurement of the absolute flux and spectrum that confirmed the excess near 5 MeV.~\cite{An:2015nua} 
The measured total flux, in terms of the IBD yield per fission, of $(5.92\pm0.14)\times 10^{-43}\ {\rm cm}^2/{\rm fission}$ was in agreement with previous reactor antineutrino measurements. 
The flux was $0.946\pm0.022$ relative to {the} prediction from the Huber-Mueller model. 
The uncertainty in the flux measurement is dominated by the knowledge of the IBD detection efficiency { and the number of protons in the target}. 

\begin{figure}[!htb]
\begin{center}
\includegraphics[height=8cm]{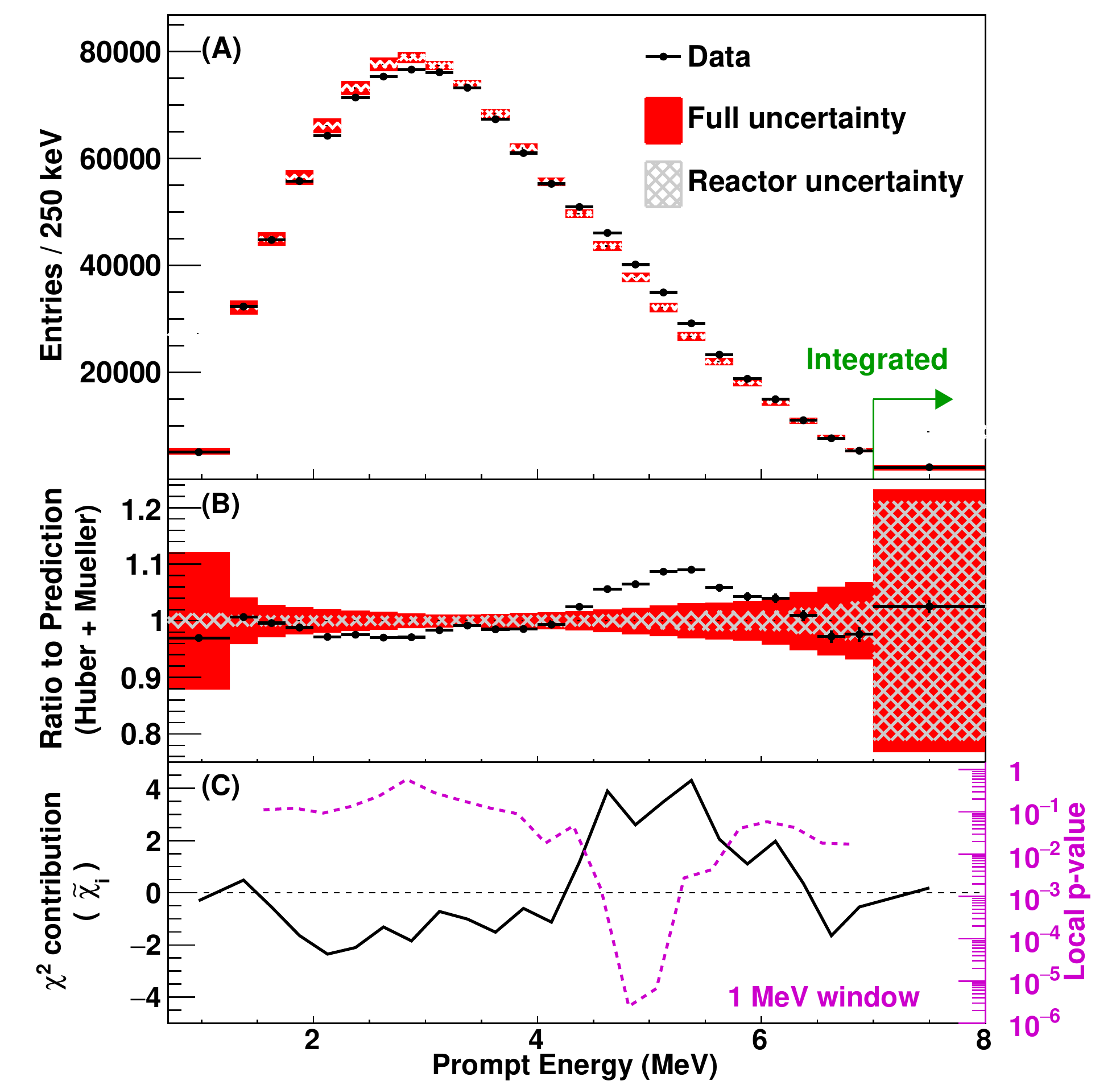}
\caption{(A) Comparison of the measured and predicted prompt spectra. 
The Huber-Mueller predicted spectrum is normalized to the number of observed IBD events. 
(B) The ratio of the measured to predicted spectrum. 
(C) The $\chi^2$ contribution to each energy bin (left axis) and the local $p$-value for a sliding 1-MeV window (right axis).
{Figure from Ref.\protect\refcite{An:2016srz}.}
}
\label{fig:shape}
\end{center}
\end{figure}

 A subsequent measurement of the spectrum with higher statistics not only observed the 5 MeV excess with high significance but also showed that the low-energy spectrum ($<\!4\ {\rm MeV}$) was not well described by the Huber-Mueller model.~\cite{An:2016srz}  
As shown in Figure~\ref{fig:shape}, the shape of the measured spectrum is inconsistent with the prediction for nearly the { entire} prompt-energy range. Reference~\refcite{An:2016srz} also provided a ``generic antineutrino spectrum'' to serve as model-independent input for reactor antineutrino experiments. 
The Daya Bay-provided spectrum was subsequently used by the NEOS experiment to set stringent limits on sterile neutrinos in the RAA region,~\cite{Ko:2016owz} and by the CONUS experiment to constrain elastic neutrino nucleus scattering.~\cite{Bonet:2020awv}

The precision on the reactor antineutrino flux was subsequently improved thanks to a comprehensive neutron calibration campaign.~ \cite{Liu:2015cra}
Detailed analysis of data and simulation reduced the uncertainty in the neutron detection component of the IBD detection efficiency by 56\%. 
The improved IBD yield was measured to be $(5.91\pm0.09)\times10^{-43}\ {\rm cm}^2/{\rm fission}$ or $0.952\pm0.014\pm0.023$ of the Huber-Mueller prediction where the first (second) uncertainty is experimental (theoretical).~\cite{Adey:2018qct}

As commercial nuclear reactors operate, the composition of the fissioning elements, known as the fuel, varies. 
Over the course of a reactor cycle, the fraction of \Ufive\ decreases from $\sim\!\!75\%$ to $\sim\!\!45\%$ while \PUnine\ (\PUone) increases from $\sim\!\!15\%$($\sim\!\!3\%$) to $\sim\!\!38\%$($\sim\!\!9\%$). The \Ueight\ fraction remains roughly constant at $\sim\!\!8\%$. 
Since both the rate and spectrum of \ANUE\ production differs for these isotopes, both the flux and spectrum of IBD events { are} expected to change during the reactor cycle. 
Previous experiments with a single \ANUE\ detector exposed to a single reactor observed flux evolution~\cite{Klimov:1994,Bowden:2008gu}. 
Daya Bay exploited these effects to determine the \ANUE\ flux produced by \Ufive\ and \PUnine. 

Daya Bay defines effective fission fractions $F_i(t)$ viewed by each AD at time $t$ as~\cite{An:2017osx}
\begin{equation}
F_i(t) = 
\sum^6_{r=1}\frac{W_{th,r}(t) \overline{p}_r f_{i,r}(t)}{L^2_r \overline{E}_r(t)} 
	\Big/ 
\sum^6_{r=1} \frac{W_{th,r}(t) \overline{p}_r }{L^2_r \overline{E}_r(t)} 
\label{eqn:efffrac}
\end{equation}
\noindent where $W_{th,r}(t)$ is the thermal power of the $r^{\rm th}$ reactor, 
$\overline{p}_r$ is the mean \ANUE\ survival probability, 
$L_r$ is the distance from the AD to the $r^{\rm th}$ reactor, 
$f_{i,r}$ is the fraction of fissions from isotope $i$ and reactor $r$,  
and 
$\overline{E}_r(t) = \sum_i f_{i,r}(t)e_i$ with $e_i$ {being} the average energy released from isotope $i$,
where $i$ runs over the isotopes \Ufive, \PUnine, \Ueight\ and \PUone. 
With this definition, the measured IBD yield per nuclear fission $\sigma_f$ is simply the sum of IBD yields from the individual isotopes, 
$\sigma_f = \sum_i F_i \sigma_i$. 
Thermal power and fission fraction data were provided on a weekly basis by the power company, validated by Daya Bay, and used to calculate effective fission fractions using Eqn.~\ref{eqn:efffrac}. 
The effective \PUnine\ fission fraction $F_{239}$ varied from about 0.23 to 0.36 for the ADs in the two near sites, EH1 and EH2, for the results in Ref.~\refcite{An:2017osx}. 
Weekly IBD datasets were grouped into eight bins in $F_{239}$ with roughly equal statistics per bin to study the effect of fuel evolution. 
The data were fitted using 
\begin{equation}
\sigma_f(F_{239}) 
= 
\bar\sigma_f
+
\frac{d\sigma_f}{dF_{239}} (F_{239} - \overline{F}_{239}) 
\label{eqn:fffit}
\end{equation}
\noindent where $\bar\sigma_f$ is the total $F_{239}$-averaged IBD yield, 
$\overline{F}_{239}$ is the average effective \PUnine\ fission fraction and 
${d\sigma_f}/{dF_{239}} $ is the change in yield per unit \PUnine\ fission fraction. 
The best fit results are 
${d\sigma_f}/{dF_{239}}  = (-1.86\pm0.18)\times 10^{-43}\ {\rm cm}^2/{\rm fission}$ 
and 
$\bar\sigma_f = (5.90\pm0.13)\times10^{-43}\ {\rm cm}^2/{\rm fission}$ 
with $\chi^2/{\rm NDF} = 3.5/6$. 
These observations differ by $3.1\sigma$ and $1.7\sigma$ from the expectations {of} the Huber-Mueller model of 
${d\sigma_f}/{dF_{239}}  = (-2.46\pm0.06)\times 10^{-43}\ {\rm cm}^2/{\rm fission}$ 
and 
$\bar\sigma_f = (6.22\pm0.14)\times10^{-43}\ {\rm cm}^2/{\rm fission}$, respectively. 
The inconsistency between data and prediction for ${d\sigma_f}/{dF_{239}} $ points to additional tension beyond the established differences in the total IBD yield $\bar\sigma_f$. 

\begin{figure}[!htb]
\begin{center}
\includegraphics[height=8cm]{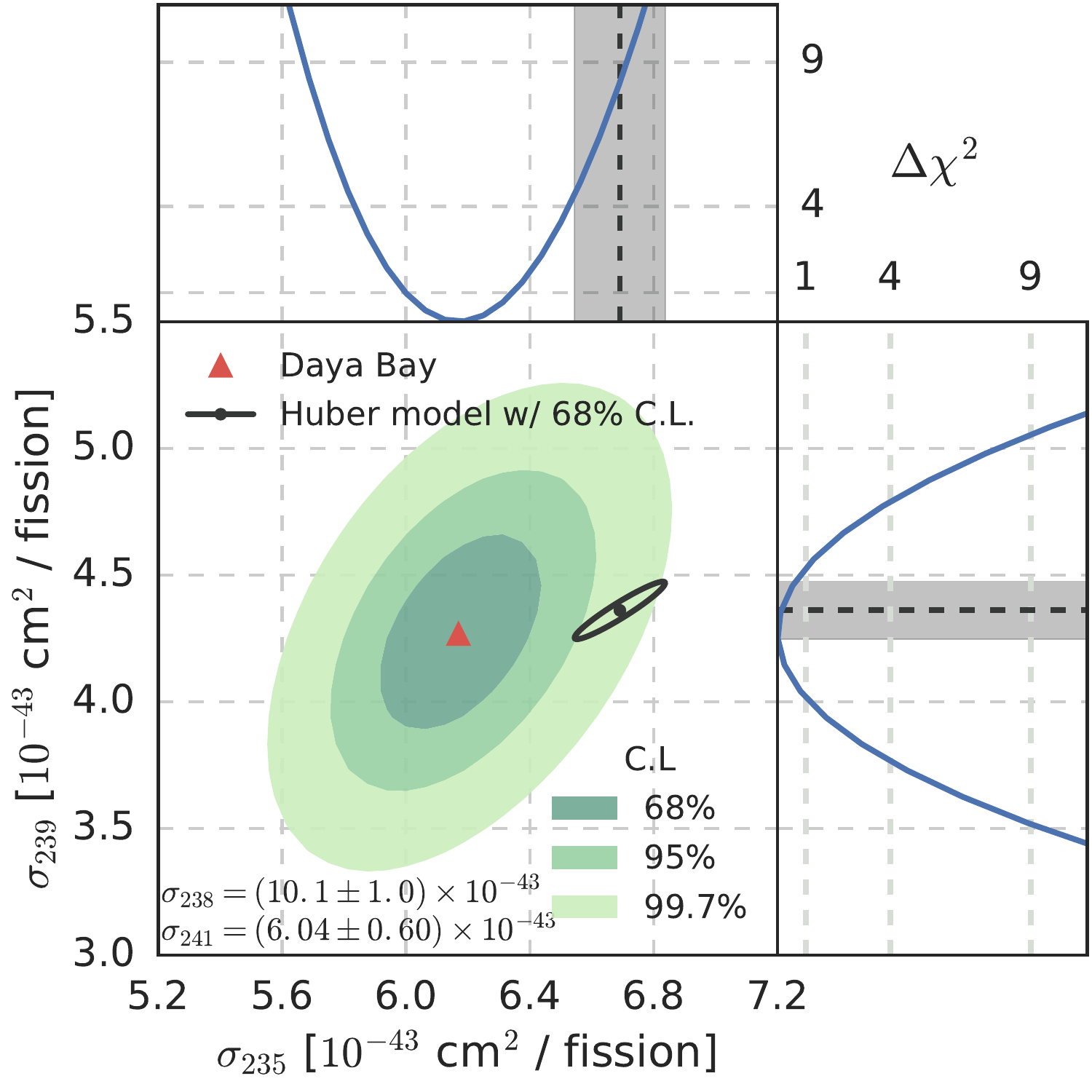}
\caption{The \Ufive\ and \PUnine\ IBD yield measurements compared with the Huber-Mueller prediction. 
The red triangle indicates the best fit $\sigma_{235}$ and $\sigma_{239}$ values and the shaded contours represent the 68\%, 95\% and 99.7\% C.L. regions. 
The assumed yields and uncertainties of $\sigma_{241}$ and $\sigma_{238}$ appear in the lower left. 
The 68\% C.L. region { of the Huber-Mueller prediction} is indicated by the black ellipse. 
The panels shows the $\Delta \chi^2$ profiles. 
{Figure from Ref.~\protect\refcite{An:2017osx}.}
} 
\label{fig:yields}
\end{center}
\end{figure}

The observed evolution was used to measure the IBD yields of \Ufive\ and \PUnine\ 
by assuming 10\% relative uncertainties on the yields on the minor fission isotopes \PUone\ and \Ueight\ taken from the Huber-Mueller model. 
The resulting IBD yields for \Ufive\ and \PUnine\ were determined to be 
$(6.17\pm0.17)$ and $(4.27\pm0.26)\times10^{-43}\ {\rm cm}^2/{\rm fission}$, respectively.
As shown in Figure~\ref{fig:yields}, the predicted and measured yields from the main fissioning isotope, \Ufive, differ significantly  
while the measurement and prediction of $\sigma_{239}$ are consistent. 
The hypothesis that all isotopes have equal fractional deficits with respect to the prediction is disfavored at $2.8\sigma$. 
This provides strong evidence that the RAA is { likely} due to the incorrect prediction of the yield from a single isotope, \Ufive, and not { solely} due to sterile neutrinos.  

There are additional experimental results that strongly disfavor sterile neutrinos as the explanation of the RAA. 
STEREO~\cite{AlmazanMolina:2020spe} recently measured $\sigma_{235} = (6.34\pm0.06[{\rm stat}]\pm0.15[{\rm syst}]\pm0.15[{\rm model}])\times10^{-43}\ {\rm cm}^2/{\rm fission}$ in good agreement with Daya Bay's measurement. 
A recent experiment measured the ratio of beta decay spectra from \Ufive\ and \PUnine\ using a technique quite different from that used to form the basis of the Huber-Mueller model.~\cite{Kopeikin:2021rnb} 
The beta decay ratio implies $\sigma_{235}/\sigma_{239} = 1.44$ according to Ref.~\refcite{Kopeikin:2021rnb} in good agreement with 1.44 determined by Daya Bay and inconsistent with 1.52 calculated by Huber-Mueller. 
(The authors of Ref.~\refcite{Kopeikin:2021rnb} claim the uncertainties in the beta decay ratio are small, but do not provide a quantitative estimate.)
Recent sterile neutrino searches by short-baseline experiments { NEOS~\cite{Ko:2016owz},}DANSS~\cite{Alekseev:2018efk}, STEREO~\cite{AlmazanMolina:2019qul} and PROSPECT~\cite{Andriamirado:2020erz} rule out the best-fit point of the RAA at { $>\!90\%$ C.L.}, $>\!95\%$ C.L., $>\!99.9\%$ C.L.  and $>\!95\%$ C.L., respectively.

These results all suggest that the calculated IBD \ANUE\ flux and spectra predicted by the Huber-Mueller model are incorrect. 
The theoretical uncertainties in the prediction are discussed in detail in Ref.~\refcite{Hayes:2016qnu} which estimates that the methodology of the Huber-Mueller prediction incur an uncertainty of 5\% that requires new experiments to resolve. 
Until such experiments are performed and analy{zed}, direct experimental measurements of the reactor {\ANUE\ }flux and spectra can provide an alternative to the models. 

\begin{figure}[!htb]
\begin{center}
\includegraphics[height=8cm]{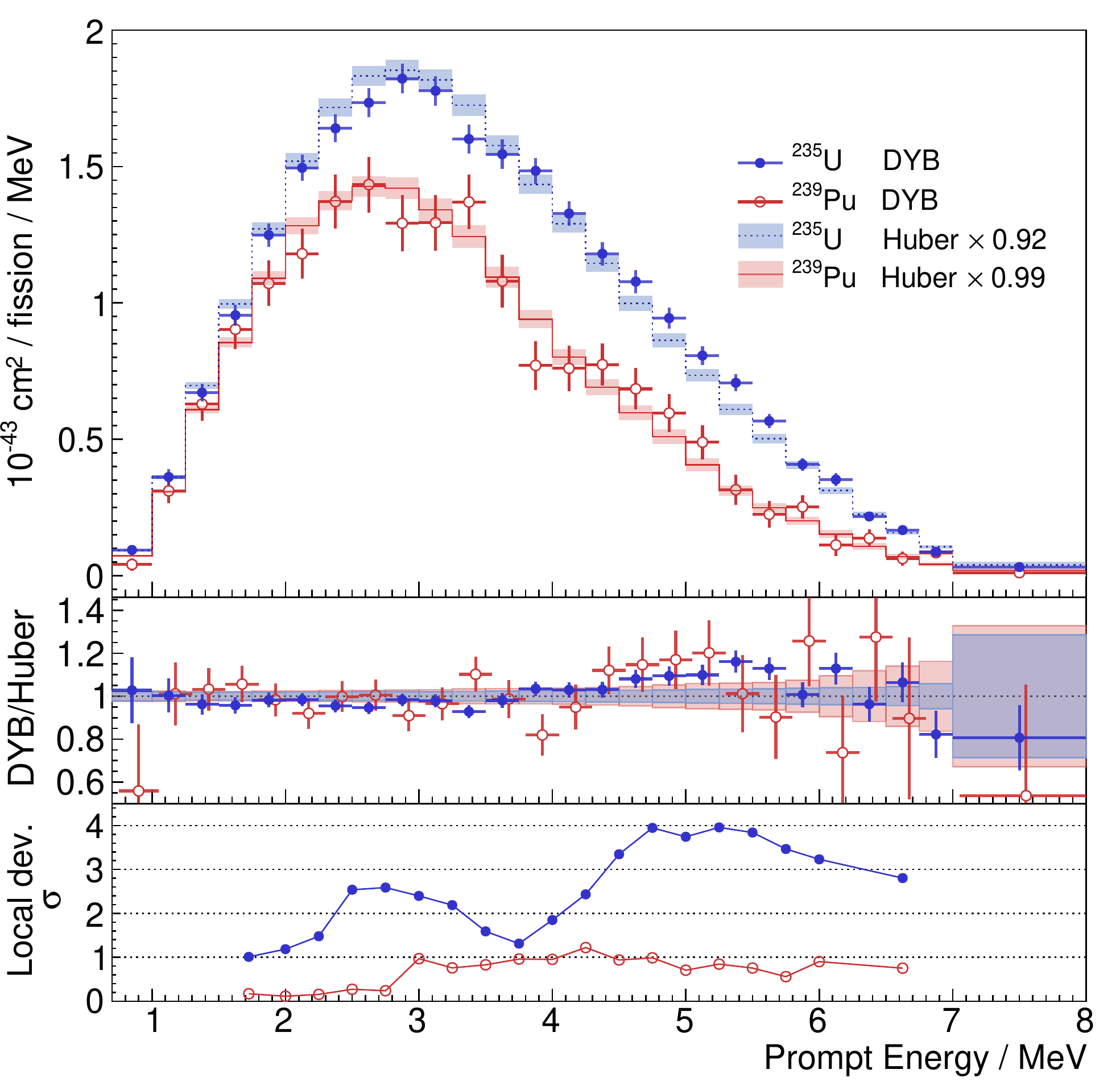}
\caption{The top panel shows the extracted \Ufive\ and \PUnine\ IBD spectra in prompt energy compared to rescaled spectra from the Huber-Mueller model. 
The middle panel shows the ratio of the spectra to the model.
The bottom panel shows the local significance in 1 MeV prompt energy windows. 
{Figure from Ref.~\protect\refcite{Adey:2019ywk}.}
} 
\label{fig:extracted}
\end{center}
\end{figure}

Daya Bay has extended the fuel evolution analysis in bins of prompt energy to extract measurements of the \ANUE\ IBD spectra from \Ufive\ and \PUnine\ fission as shown in Figure~\ref{fig:extracted}.~\cite{Adey:2019ywk}  
The resulting \Ufive\ spectrum is consistent with spectra measured at research reactors fueled exclusively by \Ufive. 
This is the first experimental measurement of the  IBD  spectrum from \PUnine\ fission daughters. 

Daya Bay { in a recent paper} deconvolves the detector response from the directly measured IBD {sample} and extract{s} \Ufive, \PUnine\  and ${\rm Pu}_{\rm combo}$ prompt-energy spectra to produce spectra in \ANUE\ energy.~\cite{Adey:2021rty} 
Here ${\rm Pu}_{\rm combo}$ is the combined spectra from \PUnine\ and \PUone\ since the evolution of these spectra are correlated. 
These \ANUE\ spectra can then be utilized to predict the IBD energy spectrum from an arbitrary reactor $A$ with fission fractions 
($f_{235}^A$, $f_{239}^A$, $f_{241}^A$, $f_{238}^A$) as
\begin{equation}
\mathbf{\cal S}^A
=
\mathbf{\cal S}_{\rm total} 
+ 
\Delta f_{235} \mathbf{\cal S}_{235}
+ 
\Delta f_{239} \mathbf{\cal S}_{\rm combo}
+ 
\Delta f_{238} \mathbf{\cal S}_{238}
+ 
(\Delta f_{241} - q \times \Delta f_{239}) \mathbf{\cal S}_{241}
\label{eqn:predictS}
\end{equation}
\noindent where $\Delta f_i \equiv f_i^A - f_i^{DB}$ where $f^{DB}_i$ is the effective fission fraction for isotope $i$ for the Daya Bay measurement, $q = 0.183$ is the ratio of \PUone\ to \PUnine\ fission fractions, 
$\mathbf{\cal S}_{\rm total}$, $\mathbf{\cal S}_{235}$ and $\mathbf{\cal S}_{\rm combo}$ are the \ANUE\ spectra from Daya Bay {while}  
$\mathbf{\cal S}_{238}$ and $\mathbf{\cal S}_{241}$ are taken from the Huber-Muller model. 
Note that the terms containing $\Delta f_i$ serve as corrections to $\mathbf{\cal S}^{\rm total}$ so that the corrections for the \Ueight\ and \PUone\ fractions are small. Ref.~\refcite{Adey:2021rty} also provides the relevant correlations between the $\mathbf{\cal S}_{i}$ to allow determination of the uncertainty on $\mathbf{\cal S}^A$. 
This technique predicts the energy spectrum to a 2\% precision for typical commercial reactors. 

\section{Other Daya Bay measurements}

 Daya Bay has also produced an independent measurement of $\theta_{13}$ using neutrons captured on hydrogen, rather than gadolinium.~\cite{An:2016bvr,An:2014ehw} 
Since the delayed energy from nH captures is 2.22 MeV rather than $\sim\!8$ MeV of nGd, more stringent criteria are required to suppress  background. 
The prompt energy is required to be greater than 1.5 MeV, compared to 0.7 MeV for the nGd analysis. 
Captures on nH are selected by requiring the delayed energy to be within $\pm3$ standard deviations of the nH capture peak and the distance between the reconstructed prompt and delayed energy deposits is required to be $<\!50$ cm. 
The coincidence time interval between the prompt and delayed is increased to [1, 400] $\mu$s due to the longer capture time in the gamma catcher region. 
Even with these more stringent criteria, the background from accidental coincidences is large. The signal-to-background rate in EH3 (EH1 and EH2) is about 1 (7). 
Based on 404 days in the eight-AD configuration combined with 217 days in the six-AD configuration, $\sin^22\theta_{13}$ is determined to be $0.071\pm0.011$, consistent with the nGd measurement. 
About half the $\sin^22\theta_{13}$ uncertainty is due to statistics and about 40\% is due to the distance criterion and the delayed-energy criterion. 

Using the same 621 day data sample, Daya Bay also searched for a time-varying \ANUE\ signal over 704 calendar days.~\cite{Adey:2018qsd}
No significant signal was observed for periods from 2 hours to nearly 2 years. 
The lack of a sidereal modulation in the time spectra was also used to limit Lorentz and CPT violating effects where Daya Bay's unique layout of multiple directions from the three reactor pairs to the three experimental halls provides unique constraints. 

Daya Bay \ANUE\ data were also examined in a model which describes the neutrino by a wave packet with relative intrinsic momentum dispersion $\sigma_{\rm rel}$~\cite{An:2016pvi}. 
Limits on $\sigma_{\rm rel}$ were derived. 
The effects due to the wave packet nature of neutrino oscillation were determined to be negligible for \ANUE\ detected in the Daya Bay experiment ensuring unbiased measurements of $\sin^22\theta_{13}$ and $\Delta m_{32}^2$ within the commonly used plane wave model. 

Two papers studied cosmogenic phenomena in Daya Bay.
The seasonal variation of the underground muon flux in the three experimental halls over a two-year period were measured.~\cite{An:2017wbm}
The results were in agreement with similar measurements by other underground experiments. 
Cosmogenic production of neutrons in the Daya Bay scintillator was also measured and compared with GEANT4~\cite{Agostinelli:2002hh} and FLUKA~\cite{Bohlen:2014buj} simulations.~\cite{An:2017jng} 
The measured production rates were 25\% to 35\% higher than the simulations in EH1 and EH2. 
The EH3 rate agreed with FLUKA's prediction but was about 30\% higher than GEANT.

\section{Predicting the future}

Daya Bay is in the process of completing final analyses using all data from 24 December 2011 to 12 December 2020. 
As described in Section~\ref{sec:osc}, with the full data sample, the relative precision on $\sin^22\theta_{13}$ and $|\Delta m_{32}^2|$ is expected to be 2.8\% and 2.1\%, respectively. 
Higher statistics measurements of the total, \Ufive\ and \PUnine\ IBD antineutrino flux and spectra should be produced. 
The higher statistics should also yield improved limits on sterile neutrinos. 
Daya Bay and PROSPECT plan to produce combined results on the \Ufive\ spectrum and sterile neutrino searches.

Many of the scientists that participated in Daya Bay are now collaborating on the JUNO experiment~\cite{Abusleme:2021zrw} that has the potential to reveal much more about the physics of neutrinos. 
I wish them luck. 

{
\section*{Acknowledgements}

I thank my Daya Bay colleagues Ming-chung Chu, Logan Lebanowski, Kam-Biu Luk, Kirk McDonald and Dmitry Naumov for comments on drafts of this article. 
I also wish to thank my BNL colleague Milind Diwan for encouraging me to write this article. 
My research is supported by  the U.S. Department of Energy under contract DE-SC0012704.

}

\end{document}